\documentclass{aa} 

\usepackage{graphicx}
\usepackage{txfonts}
\usepackage{natbib}
\usepackage{float}
\usepackage{dcolumn}
\newcolumntype{d}[1]{D{.}{.}{#1}}
\usepackage{amsmath}
\usepackage [version=4]{mhchem}

\bibpunct[]{(}{)}{;}{a}{}{,}

\usepackage{color}

\begin{document}

   \title{Millimetre and submillimetre spectroscopy of isobutene and its detection in the molecular cloud G+0.693\thanks{Transition 
          frequencies from this work as well as related data from earlier work 
          are only available in electronic form. We also provide quantum numbers, uncertainties, and residuals between 
          measured frequencies and those calculated from the final sets of 
          spectroscopic parameters. The data are available at CDS via anonymous ftp to cdsarc.cds.unistra.fr (130.79.128.5) or via https://cdsarc.cds.unistra.fr/cgi-bin/qcat?J/A+A/.}}

   \author{Mariyam Fatima\inst{1}
           \and
           Holger S.~P. M{\"u}ller\inst{1}
           \and
           Oliver Zingsheim\inst{1}
           \and
           Frank Lewen\inst{1}
           \and
           Víctor M. Rivilla\inst{2}
           \and
           Izaskun Jiménez-Serra\inst{2}
           \and
           Jesús Martín-Pintado\inst{2}
           \and 
           Stephan Schlemmer\inst{1}
           }

   \institute{$^1$I.~Physikalisches Institut, Universit{\"a}t zu K{\"o}ln,
              Z{\"u}lpicher Str. 77, 50937 K{\"o}ln, Germany\\
              $^2$Centro de Astrobiolog\'ia (CAB), CSIC-INTA, Ctra. de Ajalvir Km. 4, Torrej\'on de Ardoz, 28850 Madrid, Spain\\
              \email{fatima@ph1.uni-koeln.de and hspm@ph1.uni-koeln.de}
              }

   \date{Received xxx / Accepted xxx}
 
  \abstract
{Isobutene ((CH$_3$)$_2$C=CH$_2$) is one of the four isomers of butene (C$_4$H$_8$). Given the detection of propene (CH$_3$CH=CH$_2$) toward TMC-1, and also in the warmer environment of the solar-type protostellar system IRAS 16293$-$2422, one of the next alkenes, isobutene, is a promising candidate to be searched for in space.}
{We aim to extend the limited line lists of the main isotopologue of isobutene from the microwave to the millimetre region in order to obtain a highly precise set of rest frequencies and to facilitate its detection in the interstellar medium.}
{We investigated the rotational spectrum of isobutene in the 35$-$370~GHz range using absorption spectroscopy at room temperature. Quantum-chemical calculations were carried out to evaluate vibrational frequencies.}
{We determined new or improved spectroscopic parameters for isobutene up to a sixth-order distortion constant. These new results enabled its detection in the G+0.693 molecular cloud for the first time, where propene was also  recently found. The propene to isobutene  column density ratio was determined to be about 3:1. }
{The observed spectroscopic parameters for isobutene are sufficiently accurate that calculated transition frequencies should be reliable up to 700~GHz. This will further help in observing this alkene in other, warmer regions of the ISM. }


\authorrunning{Mariyam Fatima et al.}
\titlerunning{Rotational spectroscopy of isobutene (CH$_3$)$_2$C=CH$_2$}

\maketitle
\hyphenation{For-schungs-ge-mein-schaft}

\section{Introduction}
\label{intro}
Dense and cold molecular clouds in the interstellar medium (ISM) have been found to be a rich source of unsaturated species \citep{Brett-2017, Brett-2022}. Most of these molecules are known to have large dipole moments, such as methylcyanopolyynes CH$_{3}$C$_{2n+1}$N ($n$=1$-$4) \citep{CH3CN, CH3C3N, CH3C5N, CH3C7N}, cyanoallene CH$_{2}$CCHCN \citep{CH2CCHCN}, cyanopropenes C$_{3}$H$_{5}$CN \citep{cyanopropenes}, cyanomethanimine HNCHCN \citep{rivilla2019b}, O-bearing species such as ethenediol \citep{rivilla2022a} and \textit{n}-propanol \citep{jimenez-serra2022}, nitriles including cyanic acid (HOCN), and three C$_4$H$_3$N isomers \citep{rivilla2022c}, and propargylimine \citep{bizzocchi2020}. Some of the  detected partially or nearly saturated molecules with low dipole moments are hydrocarbons, such as propene CH$_{2}$CHCH$_{3}$ \citep{Propene_2007, Propene_2021}, butenyne CH$_{2}$CHCCH \citep{cernicharo2021_butyne}, and methylpolyynes CH$_{3}$C$_{2n}$H ($n$=1$-$3) \citep{Propyne_1981, Propyne_1981_2, Pentadiyne_1984, Methyltriacetylene_2006}. 
An up-to-date list of detected molecules can be found in the Cologne Database for Molecular Spectroscopy  (CDMS; \cite{CDMS_2016}). Only a small number of the detected molecules have low dipole moments.

The detection of saturated or nearly saturated hydrocarbons ---which have low or zero dipole moments--- is important for our understanding of the chemistry of ISM. However, such detections are challenging or impossible owing to the dipole moments of these molecules. Propene is among the largest saturated or nearly saturated hydrocarbons that have been detected not only towards dark clouds, such as the Taurus Molecular Cloud-1 (TMC-1) \citep{Propene_2007}, but also in the warmer environment of the solar-type protostellar system IRAS 16293$-$2422 \citep{Propene_2021}. An interesting open question in astrochemistry pertains to our understanding of the formation and abundance of hydrocarbons in the ISM \citep{Garcia2023}. A number of chemical models have been developed to understand their formation. Different possibilities include ion--neutral molecular reactions followed by dissociative recombination, radiative association reactions of hydrocarbons with H$_{2}$, direct hydrogenation of atoms or simple molecules on grain surfaces, and radical association in a suitable network of reactions \citep{Propene_2007, Propene_formation}.

Following propene, one of the next nearly saturated hydrocarbons, isobutene, also known as 2-methylpropene, (CH$_{3}$)$_{2}$C=CH$_{2}$, contains one C and two H atoms more than propene and is therefore a promising molecule to be searched for in the ISM. If detected, isobutene will be the second molecule with a branched carbon backbone after propyl cyanide \citep{det_i-PrCN_2014}. One of the possible formation mechanisms of isobutene could be the addition of the CH$_{3}$ radical to propene. The rotational spectrum of isobutene was studied to a limited extent in the frequency region from 8 and 35 GHz \citep{Laurie_1961, Laurie_structure_1963, DR_1975, GG_1991}. 
These works provided the rotational constants, the dipole moment value from Stark effect measurements, the internal rotation barrier height, the experimental structure from isotopic substitution, and analyses of the ground and the two lowest  torsionally excited state splittings.

In the present article, we aim to extend the existing line list of the main isotopologue of isobutene using absorption spectroscopy at room temperature. We measured transition frequencies in the range 35$-$370 GHz in order to provide a list of rest frequencies that are sufficiently accurate for radio astronomical searches not only in cold ($\sim$10~K) molecular clouds, but also in the warmer ($\gtrsim$~100~K) environments of star-forming regions (see Section 2). Further, the obtained rest frequencies of isobutene are used to facilitate its  first detection in the ISM towards the G+0.693$-$0.027 molecular cloud (G+0.693 hereafter) located in the Sagittarius (Sgr) B2 complex, close to Sgr~B2(N), which in turn is close to the Galactic Centre (Section 3.2). In this work, we also searched for other low-dipole moment hydrocarbons for which rotational line lists exist in the CDMS, and report the detection of propene and the search for 1-butyne in the G+0.693 molecular cloud (Section 3.3 and 3.4).  

\section{Laboratory spectroscopy}

\subsection{Experimental details}
\label{exptl}

\begin{figure}
\centering
\includegraphics[width=5.5cm,angle=0]{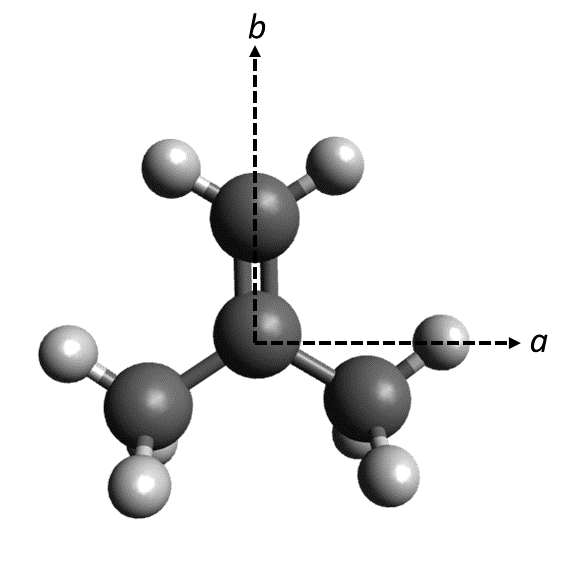}

\caption{Sketch of the isobutene molecule. The C atoms are shown in dark grey, the smaller H atoms in light grey. The $a$- and $b$-axes are shown, and the $c$-axis is perpendicular to the plane of the C atoms.}
\label{isobutene-structure}
\end{figure}

The investigation of the rotational spectrum of isobutene (Sigma-Aldrich, 99\% purity) shown in Fig.~\ref{isobutene-structure} was carried out with two different spectrometers. We employed two 7 m coupled glass cells, each with an inner diameter of  10~cm in a double-path arrangement, for measurements in the 70$-$120~GHz region, yielding an optical path length of 28~m. We used a 5~m double path cell with a 10~cm inner diameter for the 35$-$65~GHz, 170$-$250~GHz, and 250$-$370~GHz ranges. Other than the 35$-$65~GHz range, we used a tripler with in-house
developed electronics to reach the desired frequency range of 70$-$120~GHz. For the 170$-$250~GHz and 250$-$370~GHz ranges, a commercial multiplier chain (Virginia Diodes Inc., WR2.8x3 and WR4.3x2) was used. The multipliers are driven by Rohde \& Schwarz SMF~100A synthesisers as sources and Schottky diode detectors. Frequency modulation was employed  to reduce baseline effects with 2\textit{f} demodulation. This causes absorption lines to appear approximately as second derivatives of a Gaussian. Additional information on the spectrometers is available in \citet{n-BuCN_rot_2012} and \citet{OSSO_rot_2015}, respectively.

We covered the entire 70$-$120~GHz and 170$-$250~GHz ranges in steps of 500~MHz and 1~GHz, respectively. In the 35$-$65~GHz and 250$-$370~GHz ranges, individual transition frequencies were recorded in 1.5~MHz steps focusing on the well-predicted lines. The difference in recording strategy is based on the available sensitivity and line density in each frequency range. 
The pressure was set to around 2.1~Pa in the 3~mm region as test measurements showed that peak intensity was best between $\sim$2.0 and $\sim$2.5~Pa. We refilled the cells after roughly five hours to avoid pressure broadening due to small leakages. The assigned uncertainties for the newly measured lines were around 10$-$35~kHz, where a Voigt line-profile fit was used to determine the centre frequency. Most of the lines had an uncertainty of less than 20~kHz. However, when neighbouring \textit{J} transitions were overlapping, uncertainty values of 30-35~kHz were applied.


\subsection{Spectroscopic background}

Isobutene is an asymmetric rotor and with $\kappa = (2B - A - C)/(A - C) = 0.6673$ close to the oblate top limit of $+$1. Its dipole moment of 0.503~$\pm$~0.009~D is along the \textit{b}-inertial axis, which is aligned with the C=C bond; see Fig. \ref{isobutene-structure} \citep{Laurie_1961}. The molecular symmetry group of isobutene is $G_{36}$. In the rigid rotor case, the molecule has $C_{\rm 2v}$ symmetry with two equivalent methyl groups and two equivalent methylene hydrogens. \citet{Laurie_1961} and \citet{DR_1975} report the height of the threefold barrier to internal rotation $V_3$ to be around 2.21~kcal~mol$^{-1}$ and 2.170(9)~kcal~mol$^{-1}$ or $\sim$760 cm$^{-1}$, respectively, in good agreement with each other. The internal rotation of the methyl groups splits the energy levels into four substates, labelled $AA$, $AE$, $EA$, and $EE$. For consistency with previous works, we use the labels $A_1 A_1$, $A_1E$, $EA_1$, and $EE$ in this work \citep{DR_1975}, where $A_1 A_1$ is non-degenerate, $A_1E$ and $EA_1$ are doubly degenerate, and $EE$ is quadruply degenerate. Only transitions within one symmetry species, that is, only the transition corresponding to $A_1A_1$ $\gets$ $A_1A_1$ and equivalent are allowed. \citet{DR_1975} calculated that for isobutene, which has eight symmetrically interchangeable protons, each transition should be observable as a quartet with spin-weight intensity ratios of $A_1A_1 : EE : A_1E : EA_1$ , which depend on the parity of the $K_aK_c$ labels. If $K_a+K_c$ is even, the transition intensities should be 36 : 64 : 20 : 16, and if $K_a+K_c$ is odd, these should be 28 : 64 : 12 : 16. Furthermore, \citet{DR_1975} suggest that the $A_1E$ and $EA_1$ species collapse frequently when the internal rotation barrier is sufficiently high and when there is no near $K$ degeneracy. Depending on the splitting width and the instrument resolution, most of the lines of isobutene measured between 8 and 35~GHz showed singlet and triplet splitting pattern. Only in a few cases were quartet patterns observed where the $A_1E$ and $EA_1$ splitting was large enough to be resolved.


\subsection{Spectroscopic results and discussion}

\begin{figure}
\centering
\includegraphics[width=9cm,angle=0]{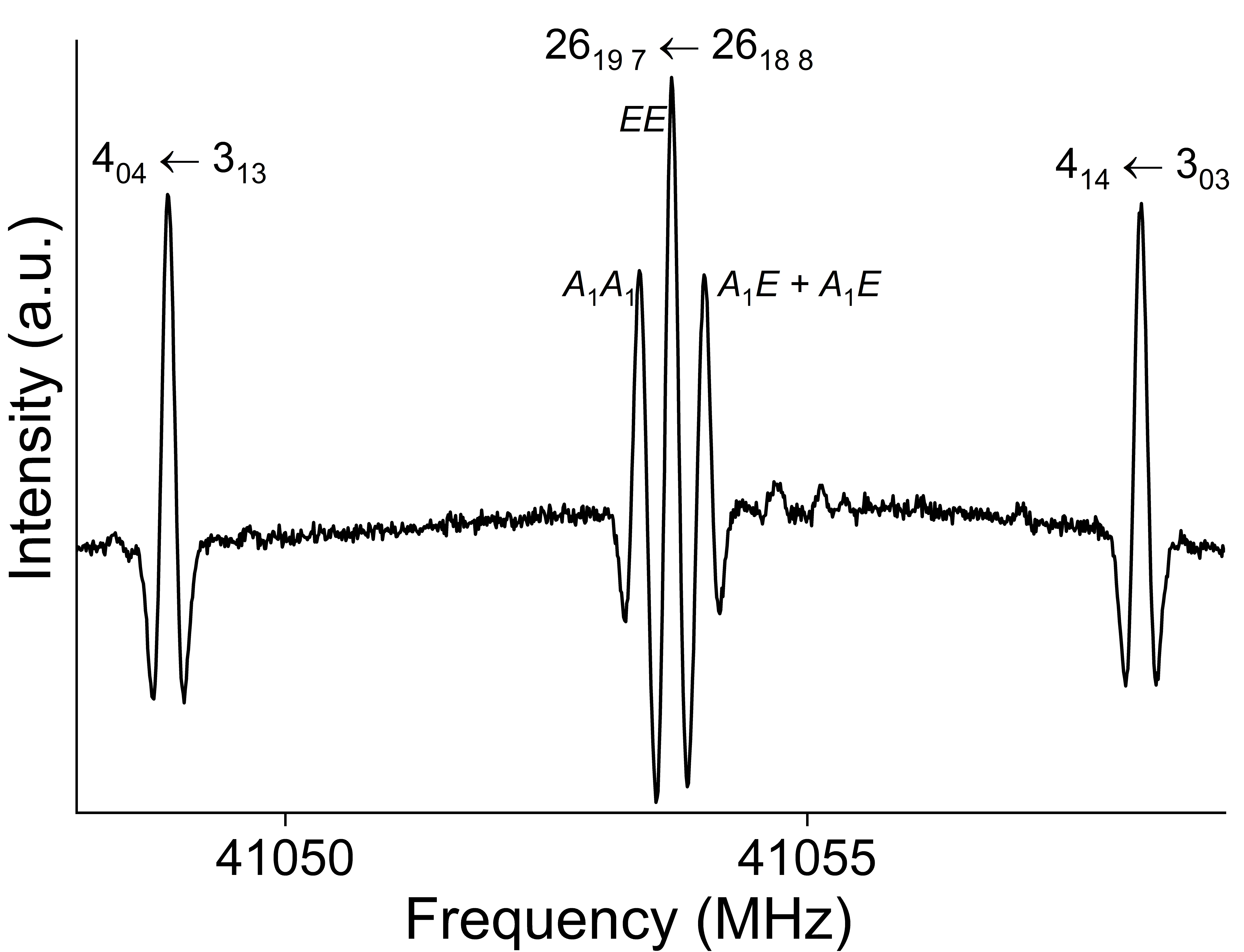}

\caption{Part of the millimetre-wave spectrum of isobutene. All transitions are split due to the internal rotation (see text). However, most of the lines are observed as unresolved single lines (as shown for $4_{0,4}\gets3_{1,3}$ and $4_{1,4}\gets3_{0,3}$ transitions) or triplets (as shown for $26_{19,7}\gets26_{18,8}$ transition) because of the high barrier to the methyl internal rotation.  The three components are assigned from left to right as $A_1 A_1$, $EE$, and unresolved $A_1E$ plus $EA_1$ with measured splitting $A_1A_1-EE$ of 0.3~MHz.}
\label{graph}
\end{figure}

We observed a \textit{b}-type spectrum of isobutene. The strongest transitions of isobutene in our data are $R$-branch transitions ($\Delta J = +1$) with $\Delta K_a = \pm 1$ and $\Delta K_c = +1$. The Boltzmann peak at 300~K is near 280~GHz. $Q$-branch transitions with $\Delta K_a = \pm 1$ and $\Delta K_c = \mp 1$ are weak but observable.

As shown in Fig.~\ref{graph}, we observed unsplit lines and triplet line patterns for the most part. The triplet transitions of isobutene are expected to be in the intensity ratio of $\sim$1~:~2~:~1. The observed fine structure due to methyl group internal rotors ---which results in the triplet pattern--- has been assigned as $A_1A_1$, $EE$, and blended $A_1E+EA_1$.  The splitting width ($A_1A_1-EE$) of the $26_{19,7} \gets 26_{18,8}$ transition shown in Fig.~\ref{graph} is 0.3~MHz. Most of the assigned $R$-branch transitions appeared as unresolved singlets, and  triplet patterns are only observed in some cases with $J \leq 17$
transitions, where the maximum splitting width is 1~MHz. Triplet patterns have been observed for $Q$-branch transitions with $J \leq 58$ transitions, with a splitting width of 5.4~MHz.

\begin{figure}
\centering
\includegraphics[width=9cm,angle=0]{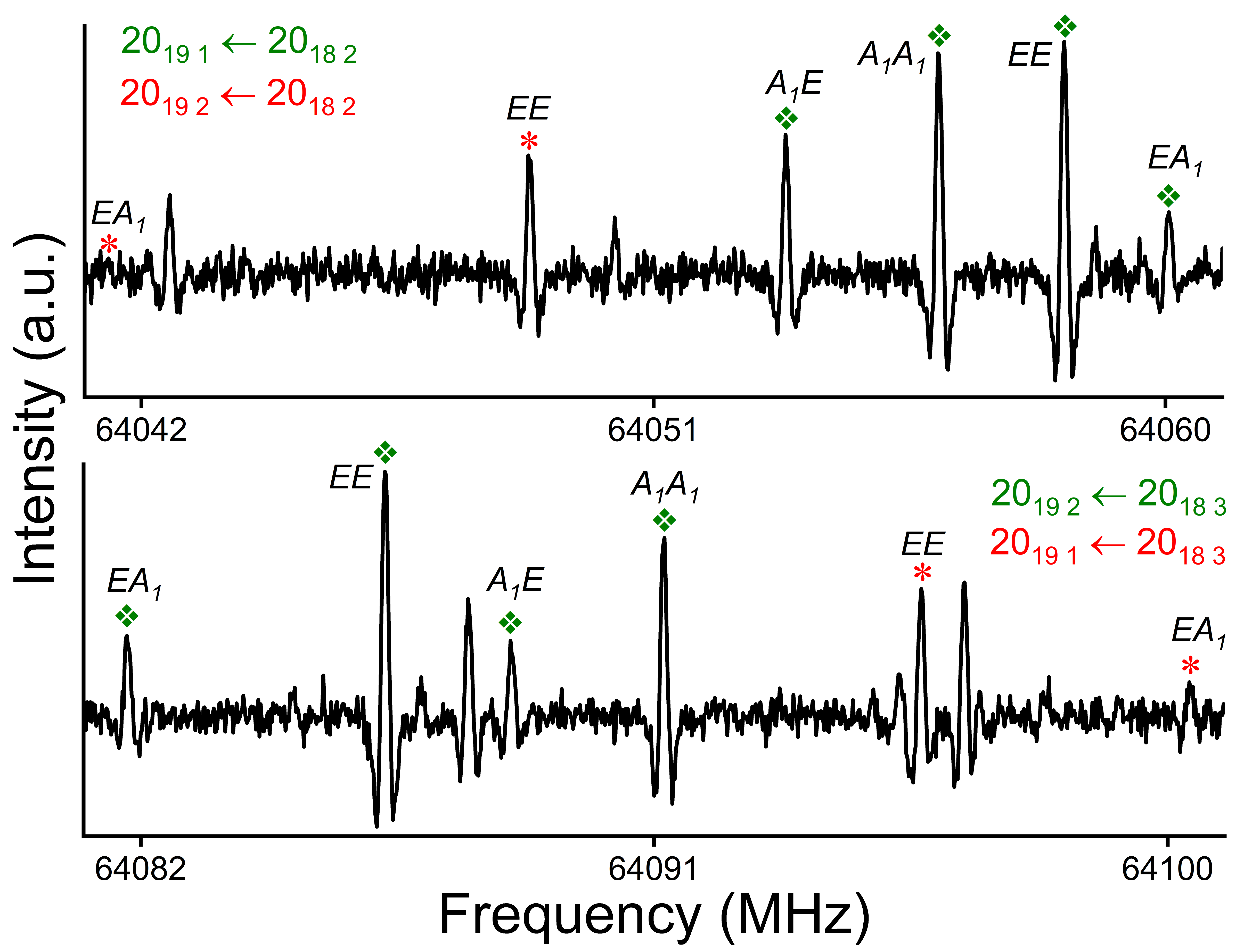}

\caption{The $20_{19,1}\gets20_{18,2}$ and $20_{19,2}\gets20_{18,3}$ transitions of isobutene highlighting the `normally' forbidden $c$-type transition resulting from the interation of the overall and internal rotation. We note that the $EA_1$ component of the $20_{19,2}\gets20_{18,2}$ was too weak to be observed.}
\label{splitting}
\end{figure}

We did not resolve the $A_1E$/$EA_1$ splitting except in a few more complex patterns in the spectrum of isobutene. As shown in Fig.~\ref{splitting}, we not only observe the allowed $20_{19,1} \gets 20_{18,2}$ $b$-type transition, but also the normally forbidden $20_{19,2} \gets 20_{18,2}$ $c$-type transition. Mixing occurs in the $A_1E$ and $EE$ characters of the internal rotation components if the internal rotation splitting and the asymmetry splitting are of similar magnitude \citep{splitting_lines, splitting-2}. This results in six lines, where four ($A_1 A_1$, $EE$, $A_1E$ and $EA_1$) belong to the former and two ($EA_1$ and $EE$) to the latter transition. The positions and intensities of each component are different from the expected pattern of $A_1 A_1$, $EE$, $A_1E,$ and $EA_1$ with intensity ratios of 36~:~64~:~20~:~16 . Similar behaviour is observed for $20_{19,2}\gets20_{18,3}$ and $20_{19,1}\gets20_{18,3}$ transitions, as shown in Fig.~\ref{splitting}. 
We observed a few additional transitions with irregular internal rotation patterns; all of them had low values of $K_c$ and $K_a$ values close to $J$.

We used the program ERHAM by \citet{erham_1997,erham_2012} to calculate and fit the rotational spectrum of isobutene. Previously assigned transitions \citep{Laurie_1961, DR_1975, GG_1991} were used for the determination of initial rotational parameters in Watson’s A reduction. 
In total, transitions up to $J = 17$ and $K_c = 4$ have been reported up to 35~GHz. Newly measured transitions from 35$-$270~GHz were added step wise to improve the data set. We started with low $J$, $K_c$ transitions and gradually included higher $J$ and $K$ transitions at higher frequencies. 
In each frequency range, the $R$-branches were added first if the predicted intensities were sufficient to be observed to get a good-quality prediction of the distortion constants. This was followed by addition of  predicted $Q$-branch transitions, which led us to introduce additional centrifugal distortion and tunnelling parameters. 
These iterations proceeded until a reasonable ERHAM fit was achieved. The new software, which is  called LLWP and is based on Loomis-Wood plots, was used to assign the transitions \citep{LLWP}. In total, the final fit includes 1176 distinct transitions.

Transitions up to $J = 60$ and $K_c = 41$ were included in the fit. A total of 19 spectroscopic parameters were determined by the least-squares method and are listed in Table \ref{tab-main-species}. The rotational parameters $A$, $B$, and $C$ are compared with the previous work from \citet{DR_1975}. 
We determined a complete set of quartic and sextic distortion parameters for the first time. In addition, two internal rotation parameters ($\rho$ and $\beta$), one energy tunnelling parameter ($\epsilon _{10}$), and one tunneling correction parameter $[(B - C)/4]_{10}$ were determined. The notation $[(B - C)/4]_{10}$ follows \citet{erham_1997} and signals a tunnelling correction to $B - C$. 
This parameter can also be viewed as an asymmetry correction to the tunnelling and could then be labelled as $\epsilon _2$. 
The weighted (dimensionless) standard deviation of the fit is 0.98, while the rms (for the unweighted frequencies) is 26.5~kHz.

The range of $J$ and $K_{c}$ quantum number and also the number of different transition frequencies for isobutene were greatly increased with respect to the previous investigations. The line list of the assigned transitions can be found in the CDMS\footnote{https://cdms.astro.uni-koeln.de/classic/predictions/daten/Isobutene/},
along with further related files.

A comparison of the observed distortion parameters with those from quantum-chemical calculations described in Appendix~\ref{vib} is not straightforward. This is due to the fact that the output is reported in the III$^{r}$ representation, which is because isobutene is an oblate rotor, whereas ERHAM employs the prolate I$^{r}$ representation only. The A reduction in an oblate representation (III$^r$ or III$^l$) has quite frequently turned out to be a poor choice,  because of the slower convergence compared to other combinations of reduction and representation; see for example \citet{2022JMoSp.38411584M}. 
The work from \citet{DR_1975} reports a limited set of spectroscopic parameters $A-C$ as well as $\Delta_{JK}$, $\Delta_{K}$, $\delta_{J}$, and $\delta_{K}$ for the ground state of isobutene from 37 $Q$-branch transitions in the III$^{r}$ representation. We fit the complete set of previous data from 8$-$35~GHz using ERHAM for a comparison with our present values. 
The results of the previous data fit can be found in Table~\ref{tab-old-data} in Appendix~\ref{previous_para}. Only the rotational constants, together with $\rho$, $\beta$, $\epsilon _{10}$, and quartic distortion parameters, were allowed to float in the fit. As the number of transitions from previous data was limited, the sextic distortion constants were not well determined in trial fits. The obtained parameters are of the same order of magnitude as those  of the present work. 
Some of the distortion parameters, $\Delta_{JK}$ and $\Delta_{J}$, were not well determined and display discrepancies compared to our present values. 
The effect of the deviations can be seen in the  rest frequencies calculated from these data, especially when going to transitions with higher $J$ values. 
For example, the transition frequency of the relatively strong $R$-branch transition $9_{0,9}\gets8_{1,8}$ is found to be $\sim$1~MHz higher than from our parameters. This difference increases further to around 12~MHz for transitions with higher $J$, as in the $20_{0,20}\gets19_{1,19}$ transition.

The $V_3$ barrier height of isobutene is known to be 760~cm$^{-1}$. However, for acetone and dimethyl ether (DME) this value is 251~cm$^{-1}$ and 903~cm$^{-1}$, respectively, which also have two methyl tops equivalent to that of isobutene \citep{Acetone,DME}. Generally, the barrier height is inversely proportional to the size of the splitting ($A_1A_1-EE$). Another factor that becomes important for methyl top molecules is $\rho$. Isobutene and acetone have similar $\rho$ values of ~0.055 and 0.06, respectively. Whereas, the  $\rho$ value
 for DME is 0.2. Therefore, the change in the splitting size with increasing $K$ is larger in DME compared to acetone and isobutene.

In the isobutene spectrum, some transitions pertaining to the two low-lying fundamental torsional states, $\varv_{15}=1$ and $\varv_{21}=1$ (Table~\ref{tab-vibration}), of isobutene were identified in our room-temperature spectra. These states were investigated previously by \citet{DR_1975}. The Boltzmann factor for these two excited states at 300 K are 0.4 and 0.36, respectively, compared to the ground state (see Table \ref{tab-vibration}). In cold sources, their population is negligible (see Table \ref{tab-partition_function}). Therefore, in the present work, we do not focus on the analysis of the torsional states as we are interested in searching for the ground state of isobutene in the ISM.

\begin{table}
\begin{center}
\caption{Present and previous experimental spectroscopic parameters (MHz) of the main isotopologue of isobutene and details of the fits.}
\label{tab-main-species}
\renewcommand{\arraystretch}{1.10}
\begin{tabular}[t]{l D{.}{.}{10} }
\hline \hline

Parameters  &   \\
\hline 
\textit{Previous work}$^{(a)}$ &  \\
\hline
$A$ /MHz                              &  9133.31(3)      \\
$B$ /MHz                              &  8381.80(3)       \\
$C$ /MHz                              &  4615.97(2)      \\
\hline
\textit{This work}$^{(b)}$ &  \\
\hline
$A$ /MHz                              & 9133.36452~(12)    \\
$B$ /MHz                              & 8381.87456~(12)    \\
$C$ /MHz                              & 4615.97690~(11)    \\
$\Delta_K$ /kHz                       & 9.09875~(12)     \\
$\Delta_{JK}$ /kHz                    & -5.46233~(12)    \\
$\Delta_J$ /kHz                       & 4.57324~(14)     \\
$\delta_K$ /kHz                       & 1.585116~(57)    \\
$\delta_J$ /kHz                       & 1.971182~(29)    \\
$\Phi_K$ /mHz                         & 104.63~(11)     \\
$\Phi_{KJ}$ /mHz                      & -101.68~(11)    \\
$\Phi_{JK}$ /mHz                      & 12.408~(74)    \\
$\Phi_J$ /mHz                         & 6.394~(60)    \\
$\phi_K$ /mHz                         & -0.960~(40)    \\
$\phi_{JK}$ /mHz                      & 14.369~(38)    \\
$\phi_J$ /mHz                         & 3.280~(10)    \\
                                      &                  \\
$\epsilon_{10}$ /MHz                  & -3.2257~(12)     \\
$[(B-C)/4]_{10}$ /kHz                 & 0.0352~(15)      \\
                                      &                  \\
$\rho \times 10^{-3}$                 & 55.8678~(23)     \\
$\beta$ / $^{\circ}$                  & 29.506~(17)      \\
                                      &                  \\
no. of lines                          & 3751             \\
no. of transitions                    & 1176             \\
standard deviation$^{(c)}$            & 0.98            \\
microwave rms /kHz                    & 26.514          \\
\hline
\end{tabular}
\end{center}
\tablefoot{
$^{(a)}$ \citet{DR_1975}\\
$^{(b)}$ Watson's A reduction in the I$^{r}$ representation is used in ERHAM. Numbers in parentheses are
one standard deviation in units of the least significant figures. The first block shows rotational and
centrifugal distortion constants, and the second and third blocks show ERHAM tunnelling parameters and
geometrical values, respectively. The bottom row contains general information
about the fits.\\
$^{(c)}$ Weighted unitless value for the respective single state fit.         
} 
\end{table}


\subsection{Considerations for generating an isobutene line list}

The search for isobutene in the ISM requires the generation of a sufficiently accurate line list. 
The variance file generated from the fit of the spectroscopic parameters to the experimental data is one important ingredient. Another fairly obvious ingredient is the permanent dipole moment, which was determined experimentally as 0.503~$\pm$~0.009~D by \citet{Laurie_1961}. An additional one is the partition function. Traditionally, this was determined only for the ground vibrational state, but in the somewhat warmer ($T \gtrsim 100$~K) and denser medium, excited vibrational states are frequently populated non-negligibly. 
The ideal solution to determine the partition function is to sum over all rovibrational states at a given temperature. However, spectroscopic parameters are rarely known extensively enough for molecules with more than three atoms. A usually good approximation is to express the total partition function $Q_{\rm tot}$ as a product of the rotational ($Q_{\rm rot}$) and the vibrational partition function ($Q_{\rm vib}$). The rotational partition function was calculated using SPCAT \citep{spfit_1991}, which sums over the ground state energy levels up to sufficiently high values of $J$ and $K_a$, which were taken as 185 and 125, respectively, for isobutene. We did not use ERHAM for this purpose because of limitations in $J$ and $K$. The reduced spin weights were employed to overcome limitations in the upper state degeneracy ($g_u$) for the catalogue form. The spin weights 36 : 64 : 20 : 16 for even states and 28 : 64 : 12 : 16 for odd states were also reduced by a factor of four for generation of $Q_{\rm rot}$. 

It is quite common that vibrational energies are not known or only known approximately, in particular for low-lying vibrational and for overtone and combination states. A usually good approximation is the harmonic approximation, and in most cases better values can be obtained if the anharmonic fundamentals are employed.


\begin{table}
\begin{center}
\caption{Rotational $Q_\mathrm{rot}$ and vibrational partition functions $Q_\mathrm{vib}$ for isobutene at temperatures (K) implemented in the CDMS.}
\label{tab-partition_function}
\renewcommand{\arraystretch}{1.3}
\begin{tabular}[t]{ D{.}{.}{3} D{.}{.}{4} D{.}{.}{3}}
\hline \hline
\multicolumn{1}{c}{Temperature (K)}    & \multicolumn{1}{c}{$Q_\mathrm{rot}$} & \multicolumn{1}{c}{$Q_\mathrm{vib}$}  \\
\hline \hline
300.000   & 1493981.0421   & 2.278     \\ 
225.000   &      970052.1240   & 1.787     \\
150.000   &      527900.3030   & 1.351     \\
75.000    &  186656.4270   & 1.046     \\
37.500    &       66044.4178   & 1.001     \\
18.750    &   23393.8153   & 1.000     \\
15.000    &       16755.3941   & 1.000     \\
9.375     &        8303.1944   & 1.000     \\
5.000     &        3256.3920   & 1.000     \\
2.725     &        1326.5615   & 1.000     \\
\hline
\end{tabular}
\end{center}
\tablefoot{
See text for details.} 
\end{table}


\begin{figure}[htbp]
\centering
\includegraphics[width=9cm,angle=0]{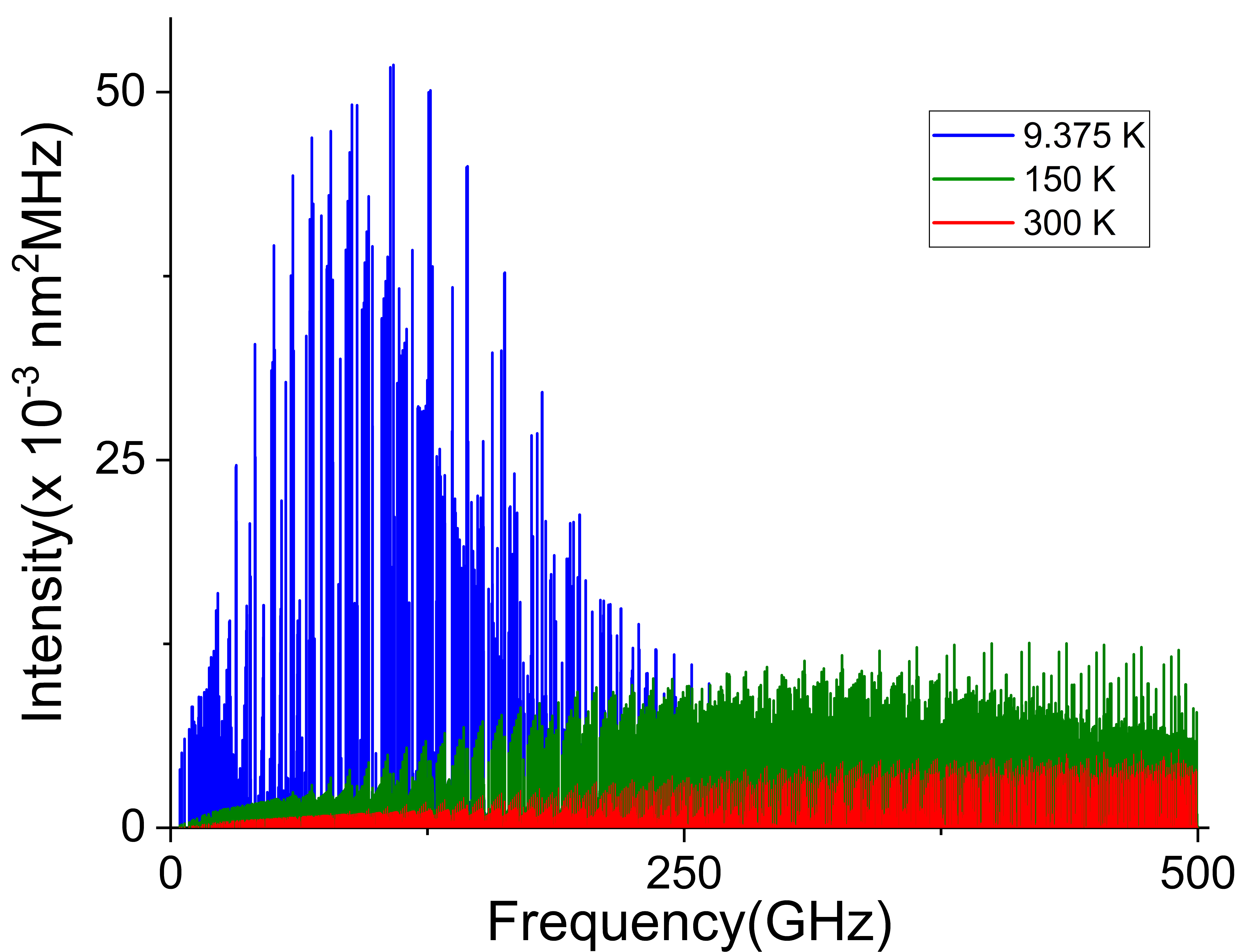}

\caption{Local thermodynamic equilibrium spectrum of the molecule, assuming two temperatures of astronomical sources where the molecule can potentially be detected: 9.375 K (dark clouds or low-excitation T regions), and 150 K (representing the environment of hot cores or corinos). For comparison, a 300 K spectrum is also shown to stress the effect of temperature on the spectrum peak. We considered the total partition function $Q_{\rm tot}$ for the simulation.}
\label{LTE}
\end{figure}

The vibrational modes of isobutene can be found in Appendix~\ref{vib}. Details of the calculation of vibrational frequencies can be found in Appendix B.
The values of $Q_{\rm rot}$ and $Q_{\rm vib}$ at temperatures commonly employed in the CDMS catalogue are given in Table~\ref{tab-partition_function}. A local thermodynamic equilibrium (LTE) spectrum simulated at 9.375~K, 150~K, and 300~K employing the corresponding $Q_{\rm tot}$ from Table~\ref{tab-partition_function} is shown in Figure \ref{LTE}. The shape of the spectrum and its peak at different temperatures can guide new observational searches.


\begin{figure*}
\centering
\includegraphics[width=19cm,angle=0]{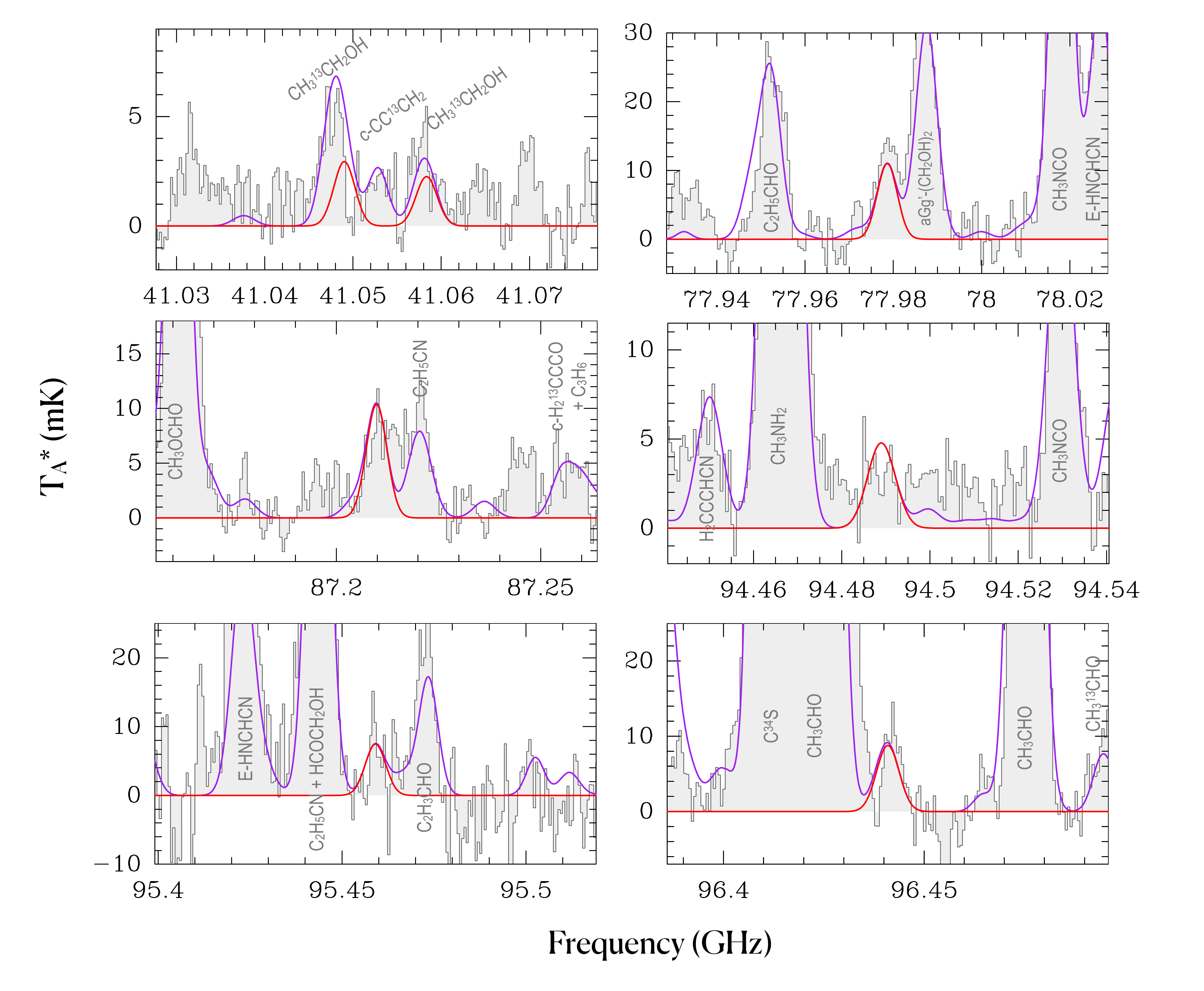}
\vspace{-0.8cm}
\caption{Unblended or slightly blended transitions of isobutene (\ce{(CH3)2C=CH_2}) detected towards the G+0.693 molecular cloud.  The black line and grey histogram show the observed spectrum. The LTE fit of isobutene is indicated with a red curve, while the emission of all the molecules identified in G+0.693, including isobutene, are indicated in purple.}
\label{fig:isobutene}
\end{figure*}

\section{Astronomical search}
\label{detection}

\subsection{Observations of the G+0.693-0.027 molecular cloud}

The molecular cloud G+0.693-0.027 (hereafter G+0.693) located in the Sgr B2 region at the centre of our Galaxy is one of the most chemically rich sources in the ISM (see e.g. \citealt{zeng2018,bizzocchi2020,rivilla2019b,rivilla2020b,rivilla2021b,rivilla2021a,rivilla2022a,rivilla2022b,rivilla2022c,rodriguez-almeida2021a,rodriguez-almeida2021b,zeng2021,jimenez-serra2022}). 
We searched for several hydrocarbons (isobutene, propene, and 1-butyne) using a high-sensitivity spectral survey towards G+0.693  carried out with the IRAM 30m telescope (Granada, Spain) and Yebes 40m telescope (Guadalajara, Spain). 
The observations were centred at $\alpha$(J2000.0)$\,$=$\,$17$^h$47$^m$22$^s$, and $\delta$(J2000.0)$\,$=$\,-\,$28$^{\circ}$21$'$27$''$. The position switching mode was used in all the observations with the off position located at $\Delta\alpha$~=~$-885$'', $\Delta\delta$~=~$290$'' from the source position.
The line intensity of the spectra was measured in units of $T_{\mathrm{A}}^{\ast}$ as the molecular emission towards G+0.693 is extended over the beam (\citealt{requena-torres_organic_2006,requena-torres_largest_2008,zeng2018,zeng2020}).
For more detailed information about the observational survey, we refer to \citet{rodriguez-almeida2021a}, \citet{rivilla2021a} and \citet{rivilla2022c}.

\begin{table*}
\centering
\caption{Derived physical parameters for the hydrocarbons analysed towards the G+0.693$-$0.027 molecular cloud. Values without uncertainties were fixed when performing the fit (see text).}
\begin{tabular}{c c  c c c c c c  }
\hline
\hline
Name & Formula & $N$   &  $T_{\rm ex}$ & $V_{\rm LSR}$ & FWHM  & Abundance$^a$ & Ref.$^b$  \\
& & ($\times$10$^{13}$ cm$^{-2}$) & (K) & (km s$^{-1}$) & (km s$^{-1}$) & ($\times$10$^{-10}$) &   \\
\hline
Propene &  \ce{CH2=CHCH3} (C$_3$H$_6$)  &  82$\pm$16 & 16$\pm$4 & 69 & 20 & 60 &  (1)  \\  
Isobutene & \ce{(CH3)2C=CH_2} (C$_4$H$_8$)  &  26$\pm$4 & 14$\pm$9 & 69 & 20 & 19 &  (1)  \\  
Propyne & \ce{CH3CCH} (\ce{C3H4}) &  170$\pm$2 & 19$\pm$1 & 69 & 20 & 126 &  (2)  \\  
1-butyne & \ce{C2H5CCH} (\ce{C4H6})  &  $<$ 2.7 & 19 & 69 & 20 & $<$ 2 &  (1)  \\ 
\hline
\hline 
\end{tabular}
\label{tab:parameters}
\vspace{0mm}
\vspace*{-1.5ex}
 \tablefoot{
 \tablefoottext{a}{We adopted $N_{\rm H_2}$=1.35$\times$10$^{23}$ cm$^{-2}$, from \citet{martin_tracing_2008}.}
 \tablefoottext{b}{References: (1) This work; (2) \citet{bizzocchi2020}.}
 }
\label{tab:g0693}
\end{table*}

\begin{figure*}
\centering
\includegraphics[width=18.5cm,angle=0]{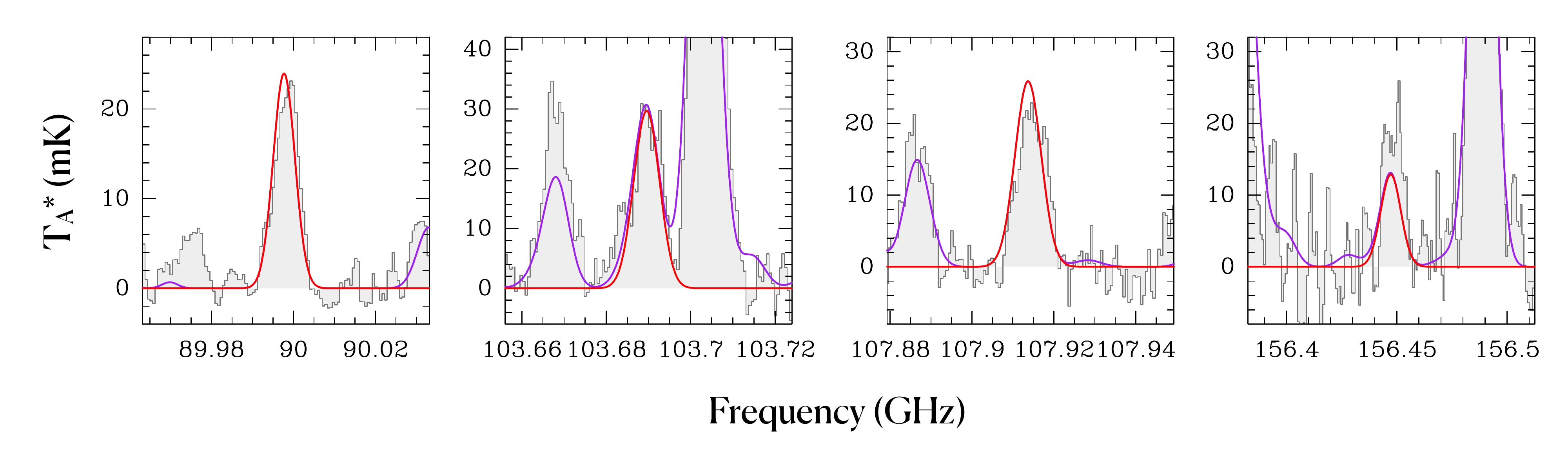}
\vspace{-0.8cm}
\caption{Selected clean transitions of propene (\ce{CH2=CHCH3}) detected towards the G+0.693 molecular cloud. The black line and grey histogram show the observed spectrum. 
The LTE fit of propene is indicated with a red curve, while the emission from all the molecules identified in G+0.693 are indicated in purple.}
\label{fig:propene}
\end{figure*}

\subsection{Detection of isobutene (\ce{(CH3)2C=CH_2})}

The Yebes 40~m observations cover a spectral range from 31.0~GHz to 50.4~GHz, while the IRAM 30~m observations cover the spectral ranges 71.77-116.72~GHz, 124.8–175.5~GHz, 199.8-222.3~GHz, and 223.3-238.3~GHz. We implemented the spectroscopic entry of isobutene from this work into the MADCUBA package{\footnote{Madrid Data Cube Analysis on ImageJ is a software developed at the Center of Astrobiology (CAB) in Madrid; http://cab.inta-csic.es/madcuba/}}; version 21/03/2023; \citealt{martin2019}).
Using the SLIM (Spectral Line Identification and Modeling) tool of MADCUBA, we generated a synthetic spectrum of isobutene under the assumption of LTE, and compare with the observed spectra.
Several groups of transitions of the molecule, which are  shown in Figure \ref{fig:isobutene}, are detected in the G+0.693 spectral survey. Their spectroscopic information is listed in Table \ref{tab:transitions}. We note that in Table \ref{tab:transitions}, the upper state degeneracy, $g_{\rm u}$, is divided by a factor of four. The reduced spin weight was employed to overcome limitation in the catalogue form. The group of transitions with $J =$ 8$-$7 at 77.978 GHz, 9$-$8 at 87.209 GHz, and 10$-$9 at 96.440 GHz appears completely unblended, without any contamination from other species. The group of transitions at 41.048 and 41.058 GHz (Table \ref{tab:transitions}) produces a doublet, which combined with emission from the $^{13}$C isotopologue of ethanol nicely reproduces the spectral profile observed (upper left panel in Fig. \ref{fig:isobutene}). The $J~=$~9$-$8 and 10$-$9 groups of transitions at 95.459 and 96.440 GHz are detected with lower signal-to-noise ratio, but they are also compatible with the observed spectra.
We note that no predicted transition by the LTE model is missing in the spectra.

We used the transitions shown in Fig. \ref{fig:isobutene} (listed in Table \ref{tab:transitions}) to derive the physical parameters of the isobutene emission. We used the SLIM-AUTOFIT tool of MADCUBA, which provides the best non-linear least-squares LTE fit to the data using the Levenberg-Marquardt algorithm. The parameters used in the LTE model are: molecular column density ($N$), excitation temperature ($T_{\rm ex}$), velocity ($V_{\rm LSR}$), and full width half maximum (FWHM) of the Gaussian line profiles. For the case of the isobutene fit, to allow the convergence of the AUTOFIT algorithm, we fixed the values of the velocity and the linewidth to $V_{\rm LSR}$=69~km~s$^{-1}$ and FWHM=20~km~s$^{-1}$, respectively, which are commonly found in other molecular tracers in G+0.693 (see \citealt{requena-torres_organic_2006,requena-torres_largest_2008,zeng2018}), and we also used these to fit other hydrocarbons, such as propyne (\citealt{bizzocchi2020}). The best fit provided by AUTOFIT, also  taking into account the emission produced by other molecules already identified in the survey, is shown with a red curve in Figure \ref{fig:isobutene}. 
We found $T_{\rm ex}$=14$\pm$7~K, and $N$=(2.6$\pm$0.4)$\times$10$^{14}$~cm$^{-2}$, which translates into a molecular abundance of 1.9$\times$10$^{-9}$ (Table \ref{tab:parameters}).

\begin{table*}
\centering
\caption{List of selected transitions of isobutene (\ce{(CH3)2C=CH_2}) and propene (\ce{CH2=CHCH3}) detected towards G+0.693, whose spectra are shown in Figures \ref{fig:isobutene} and \ref{fig:propene}, respectively. We provide the transition frequencies, quantum numbers, base 10 logarithms of the integrated intensity at 300 K (log $I$), the base 10 logarithm of Einstein coefficients (log $A_{\rm ul}$), and upper state degeneracy ($g_{\rm u}$). We note that $g_{\rm u}$ is calculated based on reduced spin weights by a factor of four. For isobutene, we have included all the hyperfine transitions that fall in the spectral ranges shown in Fig. \ref{fig:isobutene}. Some of them are  weak and their contribution to the synthetic spectrum is negligible, but we have included them for completeness.
}
\begin{tabular}{c r c c c r c l}
\hline
Species & Frequency & Transition  & log $I$ &  log  $A_{\rm ul}$  & $g_{\rm u}$ & $E_{\rm u}$ &  Blending$^b$ \\
& (GHz) & ($J'_{K_{a'},K_{c'}} \gets J"_{K_{a"},K_{c"}}(\sigma_1 \sigma_2)$)  &   (nm$^2$ MHz) & (s$^{-1}$) & & (K) &   \\
\hline
\ce{(CH3)2C=CH_2} & 41.0488473      & 4$_{0,4} -3_{1,3}(12)$ &  -7.5883 &     -7.1083 & 45   & 5.22   &  CH$_3^{13}$CH$_2$OH \\
                  & 41.0488473      & 4$_{0,4} -3_{1,3}(11)$ &  -7.4914 &     -7.1083 & 36   & 5.22   & CH$_3^{13}$CH$_2$OH \\
                  & 41.0488769      & 4$_{0,4} -3_{1,3}(01)$ &  -6.9863 &     -7.1083 & 144  & 5.22   & CH$_3^{13}$CH$_2$OH\\
                  & 41.0489065      & 4$_{0,4} -3_{1,3}(00)$ &  -7.2361 &     -7.1082 & 81   & 5.22   & CH$_3^{13}$CH$_2$OH \\
                  & 41.0581587      & 4$_{1,4} -3_{0,3}(11)$ &  -7.7130 &     -7.1080 & 36   & 5.22   & CH$_3^{13}$CH$_2$OH \\
                  & 41.0581587      & 4$_{1,4} -3_{0,3}(12)$ &  -7.5881 &     -7.1079 & 27   & 5.22   & CH$_3^{13}$CH$_2$OH \\
                  & 41.0581894      & 4$_{1,4} -3_{0,3}(01)$ &  -6.9860 &     -7.1079 & 144  & 5.22   & CH$_3^{13}$CH$_2$OH  \\
                  & 41.0582201      & 4$_{1,4} -3_{0,3}(00)$ &  -7.3450 &     -7.1079 & 63   & 5.22   &  CH$_3^{13}$CH$_2$OH \\
\hline
                 & 77.9785914       & 8$_{0,8} -7_{1,7}(12)$ &  -6.7124 &    -6.2134  & 85  & 17.54   & - \\
                 & 77.9785914       & 8$_{0,8} -7_{1,7}(11)$ & -6.6155  &    -6.2134  & 68 & 17.54    & - \\
                 & 77.9785915       & 8$_{1,8} -7_{0,7}(11)$ & -6.8373  &    -6.2133  & 51  & 17.54   & - \\
                 & 77.9785915       & 8$_{1,8} -7_{0,7}(12)$ & -6.7124  &    -6.2134  & 68  & 17.54   & - \\
                 & 77.9786193       & 8$_{0,8} -7_{1,7}(01)$ & -6.1103  &    -6.2133  & 272 & 17.54   & - \\
                 & 77.9786195       & 8$_{1,8} -7_{0,7}(01)$ & -6.1103  &    -6.2133  & 272 & 17.54   & - \\
                 & 77.9786473       & 8$_{0,8} -7_{1,7}(00)$ & -6.3602  &    -6.2133  & 153 & 17.54   & - \\
                 & 77.9786474       & 8$_{1,8} -7_{0,7}(00)$ & -6.4694  &    -6.2134  & 119 & 17.54   & - \\
\hline 
                 & 87.2098252     & 9$_{0,9} -8_{1,8}(11)$  &  -6.6912  &    -6.0612  & 57  & 21.72   & - \\
                 & 87.2098252     & 9$_{1,9} -8_{0,8}(12)$  & -6.5662   &    -6.0611  & 76  & 21.72   & - \\
                 & 87.2098252     & 9$_{0,9} -8_{1,8}(12)$  &  -6.5662  &    -6.0611  & 95  & 21.72   & - \\
                 & 87.2098252     & 9$_{1,9} -8_{0,8}(11)$  &  -6.4693  &    -6.0611  & 76  & 21.72   & - \\
                 & 87.2098526     & 9$_{1,9} -8_{0,8}(01)$  &  -5.9642  &    -6.0612  & 304 & 21.72   & - \\
                 & 87.2098526     & 9$_{0,9} -8_{1,8}(01)$  & -5.9642   &    -6.0612  & 304 & 21.72   & - \\
                 & 87.2098801     & 9$_{0,9} -8_{1,8}(00)$  & -6.3232   &    -6.0612  & 133 & 21.72   & - \\
                 & 87.2098801     & 9$_{1,9} -8_{0,8}(00)$  &  -6.2140  &    -6.0611  & 171 & 21.72   & - \\

\hline 
                & 94.4876421      & 8$_{2,6} -7_{3,5}(12)$  & -6.6924   &    -6.1025  & 85    & 23.07  & - \\
                & 94.4876421      & 8$_{2,6} -7_{3,5}(11)$  & -6.5955   &    -6.1025  & 68    & 23.07  & - \\
                & 94.4877915      & 8$_{2,6} -7_{3,5}(01)$  & -6.0904   &    -6.1026  & 272   & 23.07  &  - \\
                & 94.4879409      & 8$_{2,6} -7_{3,5}(00)$  & -6.3402   &    -6.1025  & 153   & 23.07  & - \\
                & 94.4904265      & 8$_{3,6} -7_{2,5}(11)$  & -6.8173   &    -6.1025  & 51    & 23.07  & - \\
                & 94.4904265      & 8$_{3,6} -7_{2,5}(12)$  & -6.6924   &    -6.1025  & 68    & 23.07  & - \\
                & 94.4905764      & 8$_{3,6} -7_{2,5}(01)$  & -6.0903   &    -6.1025  & 272   & 23.07  &  - \\
                & 94.4907263      & 8$_{3,6} -7_{2,5}(00)$  & -6.4494   &    -6.1025  & 119   & 23.07  & - \\
\hline 
                & 95.4590035      & 9$_{1,8} -8_{2,7}(11)$ &  -6.6748   &    -6.0010  & 57   &   25.09  & - \\
                & 95.4590035      & 9$_{1,8} -8_{2,7}(12)$ &  -6.5499   &    -6.0010  & 76   &   25.09  & - \\
                & 95.4590057      & 9$_{2,8} -8_{1,7}(12)$ &  -6.5499   &    -6.0010  & 95   &   25.09  & - \\
                & 95.4590057      & 9$_{2,8} -8_{1,7}(11)$ &  -6.4530   &    -6.0010  & 76   &   25.09  & - \\
                & 95.4590922      & 9$_{1,8} -8_{2,7}(01)$ &  -5.9478   &    -6.0010  & 304  &   25.09  & - \\
                & 95.4590944      & 9$_{2,8} -8_{1,7}(01)$ &  -5.9478   &    -6.0010  & 304  &   25.09  & - \\
                & 95.4591808      & 9$_{1,8} -8_{2,7}(00)$ &  -6.3068   &    -6.0009  & 133  &   25.09  & - \\
                & 95.4591830      & 9$_{2,8} -8_{1,7}(00)$ &  -6.1977   &    -6.0010  & 171  &   25.09  & - \\
\hline 
                & 96.4409266 & 10$_{1,10} - 9_{0,9}(11)$ & -6.5616      &    -5.9249  & 105  &   26.35  & - \\
                & 96.4409266 & 10$_{1,10} - 9_{0,9}(12)$ & -6.4366      &    -5.9249  &  84  &   26.35  & - \\
                & 96.4409266 & 10$_{0,10} - 9_{1,9}(12)$ & -6.4366      &    -5.9249  &  63  &   26.35  & - \\
                & 96.4409266 & 10$_{0,10} - 9_{1,9}(11)$ & -6.3397      &    -5.9249  &  84  &   26.35  & - \\
                & 96.4409536 & 10$_{1,10} - 9_{0,9}(01)$ & -5.8346          &    -5.9249  & 336  &   26.35  & - \\
                & 96.4409536 & 10$_{0,10} - 9_{1,9}(01)$ & -5.8346      &    -5.9249  & 336  &   26.35  & - \\
                & 96.4409805 & 10$_{1,10} - 9_{0,9}(00)$ & -6.1936      &    -5.9249  & 189  &   26.35  & - \\
                & 96.4409805 & 10$_{0,10} - 9_{1,9}(00)$ &  -6.0844     &    -5.9249  & 147  &   26.35  & - \\
\hline
\ce{CH2=CHCH3} & 89.9972320    & 5$_{1,4} - 4_{1,3}(11)$ &  -5.9650    &     -6.3190 &  44   &   14.83   & -  \\
               & 89.9981740    & 5$_{1,4} - 4_{1,3}(00)$ &  -5.9650    &     -6.3190 &  44   &   14.83   & - \\
               & 103.6891170   & 6$_{0,6} -5_{0,5}(11)$  &  -5.7493    &     -6.1109 &  52   &   17.50   & - \\
               & 103.6900200   & 6$_{0,6} -5_{0,5}(00)$  &  -5.7493    &     -6.1109 &  52   &   17.50   & - \\
               & 156.4469910   & 9$_{2,8} - 8_{2,7}(11)$ &  -5.2750    &     -5.5851 &  76   &   44.78   & - \\
               & 156.4475110   & 9$_{2,8} -8_{2,7}(00)$  &  -5.2750    &     -5.5851 &  76   &   44.78   & - \\

\hline 
\end{tabular}
\label{tab:transitions}
{\\ (a) ($\sigma_1 \sigma_2)$ refers to the ERHAM labelling of internal rotor components \citep{erham_1997},\\ where 00~=~$A_1 A_1$, 01~=~$EE$, 11~=~$A_1E$, 12~=~$EA_1$.}
\end{table*}

\subsection{Detection of propene (\ce{CH2=CHCH3})}

We also report the detection of propene towards G+0.693 (\ce{CH2=CHCH3}). The CDMS entry 42516 was used for this search. This entry is based on \citet{propene1} with most of the data coming from \citet{propene2}, and \citet{propene3}. Multiple unblended or slightly blended transitions of this species are detected in the survey, which we used to derive the physical properties of the emission. 
We show in Figure \ref{fig:propene} some selected clean transitions, whose spectroscopic information is summarised in Table~\ref{tab:transitions}. 
We use the same procedure used to fit isobutene, obtaining $T_{\rm ex}$=16$\pm$4~K, identical within uncertainties to that derived for isobutene, and $N$=(8.2$\pm$1.6)$\times$10$^{14}$~cm$^{-2}$, which translates into a molecular abundance of 6.0$\times$10$^{-9}$ (Table \ref{tab:parameters}). 
Thus, the abundance ratio between propene and isobutene, \ce{CH2=CHCH3}/\ce{(CH3)2C=CH_2}, is $\sim$3.

\subsection{Non-detection of 1-Butyne (\ce{C2H5CCH})}

Following the detection of propene and isobutene (\ce{C3H6} and \ce{C4H8} isomers, respectively), we also searched for 1-butyne, \ce{C2H5CCH} (\ce{C4H6}). 
This species was tentatively detected using a stacking technique towards the TMC-1 dark cloud by \citet{cernicharo2021_butyne}.
We employed the CDMS entry 54519 (version 1; March 2017). In this entry, the data are obtained from \citet{butyne1}, which also comprise frequencies from \citet{butyne2} and \citet{butyne3}. The species is not detected in the G+0.693 spectral survey.
We used the brightest transitions ---according to the simulated LTE model--- that are not strongly blended with brighter transitions of other molecules to derive the upper limit to the column density. 
We applied the excitation temperature, velocity, and linewidth derived for the simpler member of its family, propyne (Table \ref{tab:parameters}): $T_{\rm ex}$=19~K, $V_{\rm LSR}$=69~km~s$^{-1}$ and FWHM=20~km~s$^{-1}$. 
To derive the upper limit of the column density, we used the highest value that produces a LTE model compatible with the observed spectrum. We obtained  $N<$ 2.7$\times$10$^{13}$~cm$^{-2}$, which means an upper limit to its abundance compared to H$_2$ of $<$ 2$\times$10$^{-10}$. 
This gives a ratio of \ce{CH3CCH}/\ce{C2H5CCH} $>$ 63 which is in good agreement with that found in the dark cloud TMC-1 of $>$133. The TMC-1 ratio is calculated using the abundances derived from the tentative detection of \ce{C2H5CCH} by \citet{cernicharo2021_butyne}, and the detection of \ce{CH3CCH} (\citealt{gratier2016,cabezas2021}). 
However, this methy-/ethyl- ratio is significantly higher than those found in other families towards G+0.693, such as alcohols, thiols, aldehydes, and isocyanates, which are in the range  7$-$24 (\citealt{rodriguez-almeida2021a,rodriguez-almeida2021b,sanz-novo2022a,jimenez-serra2022}).

\section{Conclusion}
\label{conclusion}

We extended the isobutene line list to 370~GHz applying absorption spectroscopy. 
Transitions up to $J_\mathrm{max}$~=~60 and $K_\mathrm{c,max}$~=~41 were included in the fit, which led to the determination of a complete set of quartic and sextic distortion parameters for the first time. 
The rest frequencies resulting from this fit enabled us to search for isobutene in the G+0.693 molecular cloud. Within this source, we also successfully found propene and determined a propene to isobutene abundance ratio of~3. We also searched for 1-butyne.

Our final parameters for isobutene are good up to 700 GHz, and possibly higher with $J$ extending to about 70. Therefore, a search for isobutene in other
molecular clouds such as TMC-1 and the warm parts of star-forming
regions, such as in IRAS 16293$-$2422, could be promising. Molecular clouds, such as G+0.693
and TMC-1, are a rich source of atomic carbon available in the gas
phase, which drives a rich chemistry in hydrocarbon species.
Therefore, the detection of isobutene could help us to understand the
chemistry of the formation of saturated hydrocarbon species within these
molecular clouds.

Isobutene is one of four butene isomers, along with 1-butene, \textbf{syn}-2-butene, and \textbf{anti}-2-butene. The latter does not have a rotational spectrum because of its $C_{\rm i}$ symmetry, which leads to a zero dipole moment. 
Limited microwave data are available for both conformers of 1-butene \citep{1-butene_rot_1968} and for \textbf{syn}-2-butene \citep{E-2-butene_rot_1968,E-2-butene_rot_1991}. 
It will be interesting to search for these isomers in the ISM because branched molecules, such as isobutene, have so far been found to be less abundant than their unbranched isomers.  
The known examples are propyl cyanide \citep{det_i-PrCN_2014} and propanol \citep{jimenez-serra2022,det_iso-prop_2022}. 
A detection of 1-butene or \textbf{syn}-2-butene will present another pair of isomers that can be used to improve existing astrochemical models \citep{det_i-PrCN_2014,branched_astrochem_2017,det_iso-prop_2022}.


\begin{acknowledgements}
We acknowledge support by the Deutsche Forschungsgemeinschaft via the collaborative research center SFB~956 (project ID 184018867) project B3 
as well as the Ger{\"a}tezentrum SCHL~341/15-1 (``Cologne Center for Terahertz Spectroscopy''). We thank the Regionales Rechenzentrum der Universit{\"a}t zu K{\"o}ln (RRZK) for providing computing time on the DFG funded High Performance Computing System CHEOPS. V.M.R. acknowledges support from the project RYC2020-029387-I funded by MCIN/AEI /10.13039/501100011033. I.J-.S, V.M.R and J.M.-P. acknowledge funding from grants No. PID2019-105552RB-C41 and PID2022-136814NB-I00 from the Spanish Ministry of
Science and Innovation/State Agency of Research
MCIN/AEI/10.13039/501100011033 and by “ERDF A way of making Europe”.
\end{acknowledgements}

\bibliographystyle{aa}
\bibliography{isobutene-c4h8,rivilla}

\begin{appendix}
\label{appendix}

\section{Rotational parameters from previous data set}
\label{previous_para}

\citet{Laurie_1961}, \citet{DR_1975}, and \citet{GG_1991} published rotational data for isobutene; although the latter two publications did not consider earlier data in their analyses. 
We fit these data together for a comparison of the parameter values and the predictive power compared to our more extensive results.  
The data are sufficient to determine rotational and quartic centrifugal distortion parameters, which are given in Table~\ref{tab-old-data}.

\begin{table}
\begin{center}
\caption{Experimental spectroscopic parameters (MHz) of the main isotopologue of isobutene obtained by fitting all the previous available data between 8$-$35 GHz.}
\label{tab-old-data}
\renewcommand{\arraystretch}{1.10}
\begin{tabular}[t]{l D{.}{.}{10} }
\hline \hline

Parameters  &   \\
\hline \hline 
$A$ /MHz                              & 9133.3642~(13)   \\
$B$ /MHz                              & 8381.8739~(13)   \\
$C$ /MHz                              & 4615.9765~(13)    \\
$\Delta_K$ /kHz                       & 8.888~(74)    \\
$\Delta_{JK}$ /kHz                    &  -5.16~(10)    \\
$\Delta_J$ /kHz                       & 4.28~(32)     \\
$\delta_K$ /kHz                       & 1.452~(43)    \\
$\delta_J$ /kHz                       & 2.014~(15)   \\
                                      &                  \\
$\epsilon_{10}$ /MHz                  & -3.201~(24)     \\
                                      &                  \\
$\rho \times 10^{-3}$                 & 56.37~(25)     \\
$\beta$ / $^{\circ}$                  & 29.53~(10)      \\
                                      &     \\
no. of lines                          &472             \\
no. of transitions                    & 118            \\
standard deviation$^{(a)}$            & 0.89            \\
microwave rms /kHz                    &   34          \\
\hline
\end{tabular}
\end{center}
\tablefoot{
$^{(a)}$ Weighted unit-less value for the respective single state fit.         
} 
\end{table}

\newpage

\section{Vibrational spectrum of isobutene}
\label{vib}

Isobutene has 30 vibrational modes. The vibrational representation under $C_\mathrm{2v}$ symmetry is $10A_1 + 5A_2 + 6B_1 + 9B_2$. We note that the $B_1$ and $B_2$ representations are interchanged from the previous representations used in \cite{IR_Raman_1969}, \cite{Raman_1971}, and \cite{Fra_1971}. In recent representations, $B_1$ and $B_2$ accounts for vibrations along \textit{xz} and \textit{yz} planes, respectively. Therefore, we report $B_1$ and $B_2$ in their current representations.

As we do not know how reliable and how complete the reported vibrational data are, we carried out quantum-chemical calculations at the Regionales Rechenzentrum der Universit{\"a}t zu K{\"o}ln (RRZK) using the commercially available program Gaussian~09 \citep{Gaussian09E}. 
We applied the B3LYP hybrid density functional \citep{Becke_1993,LYP_1988}.  We employed the correlation consistent basis set augmented with diffuse basis functions aug-cc-pVTZ \citep{cc-pVXZ_1989}.

A list of the observed infrared and Raman frequencies in the gas and solid phases compared with calculated harmonic and anharmonic frequencies at the B3LYP level of theory is reported in Table~\ref{tab-vibration} \citep{Gordy-and-cook}. 
All 30 vibrational fundamentals were observed in solid state Raman spectra, whereas in the gas phase Raman spectrum only 22 bands were identified. 
In infrared experiments, 17 modes were observed, where in some cases \textit{PQR} band shapes were also observed, which are useful for the assignment of a vibrational mode to the symmetry class. 
The experimental frequencies are within 6\% and 4\% of the calculated harmonic and anharmonic frequencies, respectively.

The two lowest vibrational modes, $\nu_{15}$ and $\nu_{21}$, of isobutene are both Raman active, but only the $\nu_{21}$ mode is infrared active. 
Both vibrational modes were reported from Raman experiments in the solid phase \citep{Raman_1971,Fra_1971}. 
A value for $\nu_{21}$ was determined in addition in a  gas phase infrared spectroscopic study \citep{Raman_gas}. 
 Additional insight into the fundamental torsional modes of isobutene could not be extracted from the Raman spectrum in the gas phase in this last work by \citet{Raman_gas}.

The study of the torsional vibrations of isobutene were useful in determining the internal rotation barrier height. From gas phase measurements, the effective barrier was derived to be 728$\pm$3 cm$^{-1}$ (2.08$\pm$0.01~kcal~mol$^{-1}$) \citep{Raman_gas}. 
This value is similar to the microwave study, where the barrier was determined to be 2.21~kcal~mol$^{-1}$ \citep{DR_1975}. 
Interestingly, from the solid phase measurements, a higher barrier height was evaluated (979~cm$^{-1}$ or 2.8~kcal~mol$^{-1}$) \citep{Fra_1971}. 
As mentioned by \citet{DR_1975}, this difference suggests that  large intermolecular forces exist in the solid phase, which result in a higher barrier in the condensed phase. 
Therefore, this could further indicate that the torsional fundamentals in the gas phase are lower than those measured in the liquid or solid phase by Raman spectroscopy.


\begin{table*} 
\begin{center}
\caption{Quantum-chemically calculated harmonic and anharmonic vibrational frequencies (cm$^{-1}$) of isobutene and experimental data (cm$^{-1}$).}
\label{tab-vibration}
\renewcommand{\arraystretch}{1.3}
\begin{tabular}[t]{llccccc}
\hline \hline
Symmetry & Normal & \multicolumn{1}{c}{Harmonic} & \multicolumn{1}{c}{Anharmonic} & \multicolumn{1}{c}{Observed} & \multicolumn{2}{c}{Observed} \\
 &  modes & \multicolumn{1}{c}{frequency} & \multicolumn{1}{c}{frequency} & \multicolumn{1}{c}{ in Infrared} & \multicolumn{2}{c}{in Raman} \\
 \hline
  & & & & gas phase$^a$ & gas phase$^a$ & solid phase$^b$\\
    & & & & R/Q/P &  & \\
\hline \hline
$A_1$   &       $\nu_{1}$       &       3130.1  &       2987.2  &       2990/2980.4/2968        &       2989    &       2988            \\
            &   $\nu_{2}$       &       3102.4  &       2962.5  &       2940.8              &    2930    &       2942            \\
            &   $\nu_{3}$       &       3013.6  &       2911.5  &       ---                 &       2911    &       2916            \\
            &   $\nu_{4}$       &       1713.3  &       1666.6  &       1661.1              &    1655    &       1653            \\
            &   $\nu_{5}$       &       1502.4  &       1466.8  &       1469.6              &    1462    &       1458            \\
            &   $\nu_{6}$       &       1444.3  &       1413.5  &       ---                 &       1416    &       1421            \\
            &   $\nu_{7}$       &       1414.8  &       1377.4  &       1392/1381.2/1370        &       1366    &       1376            \\
            &   $\nu_{8}$       &       1085.2  &       1070.2  &       1075/1063.9/1054        &       1058    &       1058            \\
            &   $\nu_{9}$       &       814.4   &       797.5   &       ---                     &        803         &   809             \\
            &   $\nu_{10}$      &       379.3   &       393.6   &       ---                     &        383         &   389             \\
$A_2$   &       $\nu_{11}$      &       3048.1  &       2908.0  &       ---                 &       2970    &       2956            \\
            &   $\nu_{12}$      &       1471.6  &       1438.3  &       ---                 &       1459    &       1438            \\
        &       $\nu_{13}$      &       1022.2  &       998.2   &       ---                 &       ---         &   969             \\
        &       $\nu_{14}$      &       703.4   &       694.0   &       ---                 &       ---         &   685             \\
        &       $\nu_{15}$      &       168.6   &       185.2   &       ---                     &        ---         &   209$^d$         \\
$B_1$   &       $\nu_{16}$      &       3050.8  &       2909.0  &       2944.9              &    ---         &   2956            \\
            &   $\nu_{17}$      &       1489.5  &       1445.4  &       -/1443.7/1426       &    1439    &       1458            \\
            &   $\nu_{18}$      &       1108.6  &       1081.1  &       1079                &    ---         &   1085            \\
            &   $\nu_{19}$      &       923.1   &       906.8   &       906/889.7/873         &       883         &   889             \\
            &   $\nu_{20}$      &       442.3   &       445.0   &       445/429.1/412       &    431     &       438             \\
            &   $\nu_{21}$      &       210.0   &       213.9   &       197.6$^c$               &        ---     &       239$^d$         \\
$B_2$   &       $\nu_{22}$      &       3207.1  &       3064.7  &       3096/3086.0/3074        &       3079    &       3077            \\
            &   $\nu_{23}$      &       3100.8  &       2960.9  &       ---                     &        2970    &       2968            \\
            &   $\nu_{24}$      &       3008.0  &       2893.1  &       2892.9/2882             &        2892    &       2898            \\
            &   $\nu_{25}$      &       1483.4  &       1457.0  &       1458.4              &    1439    &       1430            \\
            &   $\nu_{26}$      &       1409.5  &       1377.0  &       ---                     &        1386    &       1386            \\
        &       $\nu_{27}$      &       1296.4  &       1267.3  &       1292/1281.9/1270        &       1281    &       1271            \\
            &   $\nu_{28}$      &       987.3   &       975.0   &       ---                     &        ---         &   996             \\
            &   $\nu_{29}$      &       961.2   &       951.6   &       973.7               &    972         &   958             \\
            &   $\nu_{30}$      &       439.2   &       444.1   &       ---                     &        ---         &   438             \\

\hline
\end{tabular}
\end{center}
\tablefoot{
$^{(a)}$ \citet{IR_Raman_1969}
$^{(b)}$ \citet{Raman_1971}
$^{(c)}$ \citet{Raman_gas}
$^{(d)}$ \citet{Fra_1971}
} 
\end{table*}

\end{appendix}  

\end{document}